\documentclass[aps,prd,amsmath,amssymb,nofootinbib]{revtex4}

\usepackage{graphicx}
\usepackage{xcolor}
\usepackage{dcolumn}
\usepackage{bm}
\usepackage{amsthm,physics,hyperref,mathtools,dsfont}
\usepackage[capitalise]{cleveref}
\usepackage{bbm}

\begin{document}

\title{Modular theory and affine representations on the Rindler horizon}

\author{Michele Arzano}
\email{michele.arzano@na.infn.it}

\author{Paolo Palumbo}
\email{paolo.palumbo@unina.it}

\affiliation{Dipartimento di Fisica ``E. Pancini", Universit\`a di Napoli Federico II, I-80125 Napoli, Italy}
\affiliation{INFN, Sezione di Napoli, Complesso Universitario di Monte S. Angelo,
Via Cintia Edificio 6, 80126 Napoli, Italy}

\date{\today}

\begin{abstract}
We develop a group-theoretic interpretation of the Unruh effect based on affine symmetry on a light ray and relate it to modular theory. For a massless scalar field in two spacetime dimensions inertial and uniformly accelerated observers select two different flows within the same chiral one-particle structure, respectively, null translations and dilations. Minkowski modes are adapted to translations, while Rindler modes are adapted to dilations, with the Mellin transform providing the natural bridge between them. When a Minkowski positive-frequency mode is restricted to a single Rindler wedge, its comparison with Rindler modes is non-unitary within the positive-frequency sector. Modular theory gives the corresponding operator-algebraic interpretation: on the horizon the modular flow of the half-line algebra is implemented by dilations, and the restricted vacuum satisfies the KMS condition. The affine group thus appears as the minimal symmetry structure underlying thermality on the Rindler horizon.
\end{abstract}

\maketitle

\section{Introduction}

Thermodynamic properties of causal horizons occupy a central place in our attempts to understand the interplay between spacetime geometry, quantum theory, and observer-dependent access to physical information. These effects are central to semiclassical gravity since they underlie black-hole thermodynamics \cite{Wald:1999vt,Carlip:2014pma} and the black-hole information problem \cite{Hawking:1976ra,Mathur:2008wi}, motivate thermodynamic and emergent-gravity perspectives on the Einstein equations \cite{Jacobson:1995ab,Padmanabhan:2003gd,Eling:2006aw,Parikh:2017aas,Alonso-Serrano:2020pcz}, and extend to cosmological horizons \cite{Gibbons:1977mu,Jacobson:2003wv}, with de Sitter space providing a notable example in which temperature and entropy are associated with a horizon in the absence of gravitational collapse \cite{Bousso:2002fq}.

The origin of thermality in the presence of horizons does not directly lie in ordinary statistical ignorance over microscopic degrees of freedom, but arises as a purely quantum effect due to restricting quantum correlations to a causally accessible region \cite{Su:2015oys,Higuchi:2017gcd,Arzano:2020thh,Arzano:2021cjm,Arzano:2023pnf}. In this perspective, the Unruh effect \cite{Unruh:1976db,Crispino:2007eb} offers a particularly transparent setting in which the causal restriction is fixed by the boost orbits of Minkowski
spacetime, and thus tied to symmetries, and thermality reflects the entanglement properties of quantum fields.

The Unruh effect is usually described as a mismatch between two notions of positive frequency. An inertial observer expands a quantum field in modes of positive frequency with respect to Minkowski time, whereas a uniformly accelerated observer uses Rindler time. The two decompositions are not equivalent within the positive-frequency sector, and the Minkowski vacuum is perceived by the accelerated observer as a thermal state. Although this is a standard result, its simplest origin can already be seen on the light ray that forms the Rindler horizon.

On a light ray there are two elementary geometric operations: translations and dilations of the null coordinate. These operations generate the one-dimensional affine group. The key observation is that the distinction between Minkowski and Rindler particles can be reformulated in terms of these two affine generators. Minkowski modes diagonalize null translations, while Rindler modes diagonalize dilations. Thus the two mode decompositions are not associated with unrelated structures; rather, they arise from two different spectral decompositions inside the same affine representation \cite{Arzano:2025aeo}.

This point becomes particularly transparent for a massless scalar field in two spacetime dimensions. The field separates into two independent chiral sectors, each living on a null line. Focusing on one of these sectors, the positive-frequency Minkowski modes are naturally labelled by positive light-cone momentum. The corresponding one-particle Hilbert space carries a continuous irreducible representation of the affine group. In this representation, translations act by phases in momentum space, while dilations rescale the momentum. Passing to the logarithm of the momentum turns the dilation generator into an ordinary translation generator, and the Mellin transform becomes the natural tool for diagonalizing it.

The Rindler description has an analogous structure in position space. Inside a Rindler wedge, the relevant null coordinate has a fixed sign. Taking its logarithm gives the Rindler null coordinate, and dilations of the original light-ray coordinate become translations in the Rindler coordinate. The Rindler modes are therefore dilation modes. Their frequency label is not a new representation label of the affine group; it is the spectral label of the dilation generator within the same affine representation selected by the sign of the light-cone momentum.

The crucial point is that translations along the full light ray do not preserve the positive half-line, because they can move points across the horizon. Dilations, by contrast, do preserve the half-line. Consequently, the Rindler observer retains the dilation symmetry but not the full affine symmetry as a unitary symmetry of the wedge. When one compares the Mellin dilation basis associated with Minkowski positive-frequency modes to the Rindler dilation basis on the half-line, the comparison is governed by a Gamma-function multiplier. This multiplier is not a pure phase. Its modulus contains the imbalance between positive and negative Rindler frequencies that gives the thermal factor of the Unruh effect (see also \cite{Arzano:2018oby} for an earlier discussion of the relationship between the affine algebra and thermal spectra). This provides a one-particle, representation-theoretic view of Rindler thermality. The affine group captures the minimal kinematic mechanism behind the Unruh effect: translations define inertial particles, dilations define Rindler particles, and the non-unitary comparison between the two positive-frequency splittings produces thermality.

The representation-theoretic construction developed here builds on \cite{Arzano:2025aeo}. In the present work we expand the representation-theoretic analysis given there by distinguishing more explicitly between the unitary Mellin decomposition within a fixed affine irreducible representation and the non-unitary comparison with the Rindler dilation modes on a single half-line. We then connect this one-particle picture with modular theory. The new element is the explicit identification of the same dilation flow at the two levels: at the one-particle level its spectral comparison produces the Gamma-function thermal factor, while at the algebraic level it becomes the modular flow with respect to which the restricted vacuum satisfies the KMS condition.
While modular theory establishes the modular flow and the corresponding KMS relation, it does not by itself determine the explicit comparison between the translation- and dilation-adapted one-particle descriptions. The affine analysis supplies this additional spectral information, including the Gamma-function multiplier and the origin of its asymmetry between opposite dilation eigenvalues.

The second purpose of this paper is to connect this affine picture with modular theory. In algebraic quantum field theory, the thermal nature of local observers is given by the entanglement structure of the algebras involved \cite{RevModPhys.90.045003}. In particular, the thermal nature of the vacuum restricted to a wedge is expressed by the KMS condition for the modular flow of the wedge algebra. The Bisognano-Wichmann theorem identifies this modular flow with Lorentz boosts \cite{Bisognano:1975ih}. On the horizon, Lorentz boosts reduce precisely to dilations of the null coordinate. Thus the same affine generator that labels Rindler modes also appears as the modular generator of the half-line algebra.

This relation suggests that the Unruh effect can be understood without relying on the full Poincaré structure of the ambient spacetime. Once one focuses on the horizon light ray, the essential ingredients are the half-line algebra, a suitable cyclic and separating state, positive translations, and the affine relation between translations and dilations. Under these conditions, modular theory fixes the dilation flow as the modular flow and implies the corresponding KMS property. The affine representation-theoretic argument and the modular-theoretic statement therefore describe the same phenomenon from two complementary perspectives.

The paper is organized as follows. In the next section we review the light-cone decomposition of a massless scalar field in two dimensions and its restriction to Rindler wedges. In Section III we identify the affine group acting on a horizon light ray and explain how Minkowski and Rindler modes correspond, respectively, to translation- and dilation-adapted decompositions. In Section IV we formulate the corresponding
one-particle Hilbert space as a continuous irreducible representation of the affine group and introduce the Mellin basis which diagonalizes dilations. In
Section V we compare the Mellin and Rindler dilation bases on the half-line and show how their non-unitary spectral comparison contains the thermal factor of
the Unruh effect. Finally, in Section VI we reinterpret the same structure in modular theory, where the dilation flow appears as the modular flow of the half-line
algebra and the vacuum restricted to the wedge satisfies the KMS condition. The last section is devoted to a summary and outlook for future work.

\section{Light-cone Minkowski and Rindler fields in two spacetime dimensions}

We begin by recalling the elementary light-cone and Rindler decompositions of the two-dimensional massless scalar field, since these decompositions provide
the concrete mode bases whose affine interpretation will be developed in the following sections. The dynamics of a real massless scalar field \(\phi(t,x)\) in two-dimensional Minkowski spacetime is governed by the Klein-Gordon equation
\begin{equation}
(\partial_t^2-\partial_x^2)\phi(t,x)=0,
\label{wave equation}
\end{equation}
whose solutions can be expanded in Minkowski plane waves as
\begin{equation}
\phi(t,x)=\frac{1}{\sqrt{4\pi}}\int_{-\infty}^{+\infty}\frac{dk}{\sqrt{|k|}}
\left[a(k)e^{i(kx-|k|t)}+a^\dagger(k)e^{-i(kx-|k|t)}\right].
\label{phi iniz}
\end{equation}
The coefficients \(a(k)\) and \(a^\dagger(k)\) are, at this stage, simply mode coefficients; the notation is chosen so as to anticipate their interpretation as annihilation and creation operators upon quantization. Let us introduce null coordinates
\[
u=x-t,
\qquad
v=x+t.
\]
In these variables, the field equation takes the form
\[
\partial_u\partial_v\phi(u,v)=0
\]
so its general solution separates into a sum of independent right-moving and a left-moving contributions
\begin{equation}
\phi(u,v)=\chi(u)+\psi(v).
\label{chi psi}
\end{equation}
This decomposition makes explicit the chiral structure of the massless theory in two dimensions: the two sectors evolve independently and may be analyzed separately. The explicit form of the two components can be read directly from the mode expansion \eqref{phi iniz}. One finds
\begin{equation}
\chi(u):=\frac{1}{\sqrt{4\pi}}\int_0^{+\infty}\frac{dk}{\sqrt{k}}
\left[a(k)e^{iku}+a^\dagger(k)e^{-iku}\right],
\label{chi u}
\end{equation}
and
\begin{equation}
\psi(v):=\frac{1}{\sqrt{4\pi}}\int_0^{+\infty}\frac{dk}{\sqrt{k}}
\left[a(-k)e^{-ikv}+a^\dagger(-k)e^{ikv}\right].
\label{psi v}
\end{equation}
It is useful to further split each chiral sector according to the sign of its null coordinate. Using the Heaviside step function, we write
\begin{equation}
\psi(v)=\Theta(v)\psi_+(v)+\Theta(-v)\psi_-(v)
\label{psi hvs}
\end{equation}
and
\begin{equation}
\chi(u)=\Theta(u)\chi_+(u)+\Theta(-u)\chi_-(u).
\label{chi hvs}
\end{equation}
The null lines \(u=0\) and \(v=0\) thus partition the full solution into four pieces, corresponding to the four sign combinations of \(u\) and \(v\). 

This decomposition is especially useful when discussing uniformly accelerated observers. Their worldlines are hyperbolae in Minkowski spacetime, and are conveniently parametrized by Rindler coordinates. In the right Rindler wedge,
\[
x>|t|,
\]
one introduces coordinates \((\eta,\chi)\) through
\[
t=\frac{e^{\alpha\chi}}{\alpha}\sinh(\alpha\eta),
\qquad
x=\frac{e^{\alpha\chi}}{\alpha}\cosh(\alpha\eta),
\]
with \(-\infty<\eta,\chi<+\infty\). The constant \(\alpha\) has dimensions of inverse length and equals the proper acceleration of the observer whose trajectory sits at \(\chi=0\). An analogous construction applies in the left Rindler wedge,
\[
x<-|t|.
\]
In null coordinates, the two wedges are characterized by
\[
\textbf{R}:\quad u>0,\quad v>0,
\qquad
\textbf{L}:\quad u<0,\quad v<0.
\]
Their boundaries are precisely the null horizons \(u=0\) and \(v=0\). From this perspective, restricting the field to a given wedge amounts to selecting the appropriate sectors in the decomposition \eqref{psi hvs} and \eqref{chi hvs}.

To be definite, let us focus on the right wedge and on the \(v\)-dependent sector. There the relevant contribution is \(\psi_+(v)\), namely
\begin{equation}
\psi_+(v)=\frac{1}{\sqrt{4\pi}}\int_0^{+\infty}\frac{dk}{\sqrt{k}}
\left[a(-k)e^{-ikv}+a^\dagger(-k)e^{ikv}\right],
\qquad v>0.
\label{psi +}
\end{equation}
Since \(v\) is positive in the right wedge, it is natural to introduce the Rindler null coordinate
\[
\xi:=\chi+\eta=\frac{1}{\alpha}\log(\alpha v).
\]
Setting \(\alpha=1\) for simplicity, the positive-frequency Rindler modes are powers of \(v\),
\[
e^{-i\omega\xi}=v^{-i\omega},
\]
with \(\omega>0\). The field \(\psi_+(v)\) may then be expanded in this basis as


\begin{equation}
\psi_+(v)=\frac{1}{\sqrt{4\pi}}\int_0^{+\infty}\frac{d\omega}{\sqrt{\omega}}
\left[v^{-i\omega}B(\omega)+v^{i\omega}B^\dagger(\omega)\right].
\label{Rind pla exp}
\end{equation}
Here \(B(\omega)\) and \(B^\dagger(\omega)\) are the coefficients of the Rindler-mode expansion. As in the Minkowski case, the notation is chosen to anticipate the operator interpretation that emerges after quantization. The comparison between the Minkowski expansion in plane waves and the Rindler expansion in powers of the null coordinate already suggests the central structural point: the former is adapted to translations along the light ray,
whereas the latter is adapted to dilations. We now make this statement precise by identifying the affine group acting on the null coordinate.

\section{Affine interpretation of the light-cone expansion}
\label{sec:affine-interpretation-light-cone}

The preceding light-cone and Rindler expansions have a natural representation-theoretic interpretation. On a fixed null ray, for definiteness the \(v\)-ray, the relevant transformations are translations and dilations,
\begin{equation}
v\mapsto v+\gamma,
\qquad
v\mapsto e^{\lambda}v .
\end{equation}
These transformations have a spacetime interpretation: the translation of \(v\) is a null translation generated by $P_v=i\partial_v=\frac{i}{2}(\partial_t+\partial_x)$
while the generator of dilations of \(v\)
\[
R=-i\,v\partial_v,
\]
is the restriction to a \(v\)-ray of the Lorentz boost generator
\[
K=-i(x\partial_t+t\partial_x)
=-i\left(v\partial_v-u\partial_u\right).
\]
These transformations comprise the one-dimensional affine group, or \(ax+b\) group,
\[
G=\mathbb R\rtimes \mathbb R_+ .
\]
Equivalently, in the terminology of \cite{Moses:1974ez}, this is the one-dimensional scale-translation group. Their construction shows that physical quantities may be decomposed into coefficients transforming under irreducible unitary representations of this group. For a scalar function \(F(v)\), the action of translations $\mathsf T(\gamma)$ and dilations $\mathsf S(\lambda)$ is given by
\begin{equation}
(\mathsf T(\gamma)F)(v)=F(v+\gamma),
\qquad
(\mathsf S(\lambda)F)(v)=F(e^{-\lambda}v),
\label{eq:affine-action-null}
\end{equation}
so that
\begin{equation}
\mathsf S(\lambda)\mathsf T(\gamma)\mathsf S(-\lambda)
=
\mathsf T(e^\lambda\gamma).
\label{eq:affine-null-law}
\end{equation}

More generally, the dilation law of a physical quantity may contain a multiplicative factor determined by its scaling weight. In the realization considered here this weight vanishes, since a massless scalar field in two dimensions has zero canonical dimension; dilations therefore act only by rescaling its argument, as in Eq.~\eqref{eq:affine-action-null}. Consider now the positive-Minkowski-frequency part of the right-moving \(v\)-sector,
\begin{equation}
\psi^{(+)}(v)
=
\frac{1}{\sqrt{4\pi}}
\int_0^\infty\frac{dk}{\sqrt{k}}\,
a(-k)e^{-ikv}.
\label{eq:psi-positive-affine-bridge}
\end{equation}
The plane waves \(e^{-ikv}\) diagonalize the translation subgroup. Thus the Minkowski light-cone expansion is translation-adapted. After the redefinition of the coefficient
\begin{equation}
\Psi(k)\propto k^{1/2}a(-k),
\end{equation}
the coefficient \(\Psi(k)\) belongs naturally to
\[
L^2\!\left(\mathbb R_+,\frac{dk}{k}\right),
\]
and the affine group acts as
\begin{equation}
\bigl(U(\gamma,\lambda)\Psi\bigr)(k)
=
e^{-i\gamma k}\Psi(e^\lambda k).
\label{eq:affine-irrep-bridge}
\end{equation}
This is the positive continuous irreducible representation of the affine group, which we will discuss in detail in the next section. The opposite sign of the light-cone momentum gives the negative continuous irreducible representation. For a real field, the positive- and negative-frequency parts are related by complex conjugation, as usual.

The Rindler expansion of the light cone field corresponds to a different diagonalization within the same representation-theoretic structure. In the right wedge \(v>0\), the
logarithmic coordinate
\begin{equation}
\xi=\log v
\end{equation}
turns dilations of \(v\) into translations of \(\xi\). The Rindler modes
\begin{equation}
e^{-i\omega\xi}=v^{-i\omega}
\end{equation}
therefore diagonalize the dilation generator rather than the translation generator. Hence the label \(\omega\) is not a label of a new affine irreducible representation. It is the spectral parameter of the dilation generator inside the
fixed affine irrep selected by the sign of the light-cone momentum.

The following section makes this statement precise by constructing the one-particle Hilbert space on the positive momentum orbit \(k>0\), exhibiting the affine invariant measure \(dk/k\), and identifying the Mellin transform as
the unitary change of basis from translation-adapted light-cone modes to dilation-adapted modes.

\section{The light-cone particle as an affine irreducible representation}
\label{sec:one-particle-affine}

We now turn the affine interpretation of the previous section into an explicit one-particle Hilbert space construction. In the right-wedge \(v\)-sector discussed above, the positive-Minkowski-frequency
part is expanded in modes \(e^{-ikv}\), \(k>0\). The sign convention in the translation phase below is chosen to match this plane-wave convention. Given $G=\mathbb R\rtimes \mathbb R_+$, we parametrize a group element by \((\gamma,\lambda)\), where \(\gamma\in\mathbb R\) is a translation parameter and \(e^\lambda>0\) is a dilation. The group law is
\begin{equation}
(\gamma,\lambda)(\gamma',\lambda')
=
(\gamma+e^\lambda\gamma',\,\lambda+\lambda'),
\label{eq:affine-group-law}
\end{equation}
which is equivalent to \eqref{eq:affine-null-law}. 

Before introducing the affine representation, it is useful to clarify the relation between the Hilbert spaces used below. The momentum amplitudes appearing in the positive-frequency field expansion are naturally described by functions
\(f(k)\in L^2(\mathbb R_+,dk)\). The affine-invariant measure is instead \(dk/k\). The two realizations are unitarily equivalent under the change of normalization

\begin{equation}
f(k)\longmapsto\psi(k)=k^{1/2}f(k),
\end{equation}
since
\begin{equation}
\int_0^\infty\frac{dk}{k}\, |\psi(k)|^2
=\int_0^\infty\, dk|f(k)|^2.
\end{equation}

We shall use \(L^2(\mathbb R_+,dk/k)\) for the affine representation. Its Mellin spectral decomposition will subsequently be realized on \(L^2(\mathbb R,d\omega)\), while the right-half-line profiles introduced in Sec.~IV.B belong to \(L^2(\mathbb R,d\xi)\). These spaces therefore describe, respectively, the affine momentum representation, its dilation spectral representation, and the Rindler half-line realization.

Let us focus on the positive light-cone momentum orbit \(k>0\). The corresponding continuous affine irreducible representation is realized on
\begin{equation}
\mathcal H_{\rm aff}^{(+)}
=
L^2\!\left(\mathbb R_+,\frac{dk}{k}\right),
\label{eq:Haff-revised}
\end{equation}
with action
\begin{equation}
\bigl(U(\gamma,\lambda)\psi\bigr)(k)
=
e^{-i\gamma k}\,\psi(e^\lambda k)\,
\label{eq:U-action-revised}
\end{equation}
which corresponds with \eqref{eq:affine-irrep-bridge} in the previous section. The measure \(dk/k\) is invariant under \(k\mapsto e^\lambda k\), and therefore dilations act unitarily:
\begin{equation}
\int_0^\infty \frac{dk}{k}\,
|\psi(e^\lambda k)|^2
=
\int_0^\infty \frac{dk}{k}\,
|\psi(k)|^2.
\end{equation}
Thus the affine invariant inner product is
\begin{equation}
\langle\psi,\phi\rangle_{\rm aff}
=
\int_0^\infty\frac{dk}{k}\,
\psi^*(k)\phi(k).
\label{eq:aff-inner-revised}
\end{equation}
This is the dilation-adapted realization of the inner product. If one instead works with translation-normalized wave packets in \(L^2(\mathbb R_+,dk)\), the unitarily equivalent affine coefficient is obtained by the map
\begin{equation}
f(k)\in L^2(\mathbb R_+,dk)
\quad\longmapsto\quad
\psi(k)=k^{1/2}f(k)\in L^2(\mathbb R_+,dk/k).
\label{eq:half-density-map}
\end{equation}

Writing
\[
\mathsf T(\gamma)=e^{-i\gamma P},
\qquad
\mathsf S(\lambda)=e^{-i\lambda R},
\]
one obtains
\begin{equation}
(P\psi)(k)=k\psi(k),
\qquad
(R\psi)(k)=i\,k\frac{d\psi}{dk}(k),
\label{eq:P-R-k-realization}
\end{equation}
and hence
\begin{equation}
[R,P]=iP.
\label{eq:affine-algebra-revised}
\end{equation}
We can introduce the momentum kets \(|k\rangle_+\) associated to the irreps \eqref{eq:Haff-revised} normalized by
\begin{equation}
{}_+\langle k|k'\rangle_+
=
k\,\delta(k-k'),
\qquad
\mathbf 1_+
=
\int_0^\infty \frac{dk}{k}\,
|k\rangle_+\,{}_+\langle k|.
\label{eq:k-normalization-revised}
\end{equation}
The action of finite transformations on the kets is given by
\begin{equation}
\mathsf T(\gamma)|k\rangle_+=e^{-i\gamma k}|k\rangle_+,
\qquad
\mathsf S(\lambda)|k\rangle_+=|e^{-\lambda}k\rangle_+.
\label{eq:kets-transform-revised}
\end{equation}
While for the generator of translations we have 
\begin{equation}
    P|k\rangle_+=k|k\rangle_+,
\end{equation}
differentiating the finite dilation action
\[
\mathsf S(\lambda)|k\rangle_+=|e^{-\lambda}k\rangle_+
\]
and using \(\mathsf S(\lambda)=e^{-i\lambda R}\), one obtains the action of the dilation generator on momentum,
\[
R|k\rangle_+
=
-i\,k\frac{\partial}{\partial k}|k\rangle_+ .
\]
Thus the \(|k\rangle_+\) diagonalize \(P\), while \(R\) moves one along the positive momentum orbit.

\subsection{The Mellin basis}
\label{subsec:mellin-basis}

The dilation generator becomes an ordinary derivative by passing to the logarithmic momentum variable
\begin{equation}
    y=\log(k/\rho),
\end{equation}
where \(\rho\) is an arbitrary reference inverse length which fixes the origin of \(y\). In terms of \(y\),
\[
\frac{dk}{k}=dy,
\qquad
R=i\frac{d}{dy},
\]
and the eigenfunctions of \(R\) are therefore
\begin{equation}
\langle k|\omega\rangle_{\rm M}
=
\frac{1}{\sqrt{2\pi}}
\left(\frac{k}{\rho}\right)^{-i\omega},
\qquad
R|\omega\rangle_{\rm M}
=
\omega|\omega\rangle_{\rm M},
\label{eq:mellin-eigenfunctions-revised}
\end{equation}
obeying 
\begin{equation}
{}_{\rm M}\langle\omega|\omega'\rangle_{\rm M}
=
\delta(\omega-\omega'),
\qquad
\int_{-\infty}^{+\infty}d\omega\,
|\omega\rangle_{\rm M}{}_{\rm M}\langle\omega|
=
\mathbf 1_+ .
\label{eq:mellin-orth-revised}
\end{equation}
With this convention the Mellin transform is given by (see Appendix B)
\begin{equation}
\psi_{\rm M}(\omega)
=
{}_{\rm M}\langle\omega|\psi\rangle
=
\frac{1}{\sqrt{2\pi}}
\int_0^\infty\frac{dk}{k}
\left(\frac{k}{\rho}\right)^{i\omega}
\psi(k),
\label{eq:Mellin-transform-revised}
\end{equation}
with inverse
\begin{equation}
\psi(k)
=
\frac{1}{\sqrt{2\pi}}
\int_{-\infty}^{+\infty}d\omega\,
\left(\frac{k}{\rho}\right)^{-i\omega}
\psi_{\rm M}(\omega).
\label{eq:Mellin-inverse-revised}
\end{equation}
The Mellin transform is a unitary map since in the variable \(y=\log(k/\rho)\) it is just the Fourier transform. In fact
\begin{equation}
\int_{-\infty}^{+\infty}d\omega\,
|\psi_{\rm M}(\omega)|^2
=
\int_0^\infty\frac{dk}{k}\,
|\psi(k)|^2.
\label{eq:Mellin-unitary-revised}
\end{equation}
Thus the label \(\omega\) is not a new affine-irrep label but it is the spectral label of the dilation generator inside one fixed affine irrep. In this basis dilations are diagonal:
\begin{equation}
\bigl(\mathcal M\mathsf S(\lambda)\psi\bigr)(\omega)
=
e^{-i\lambda\omega}\psi_{\rm M}(\omega).
\label{eq:dilation-diagonal}
\end{equation}
By contrast, the translation generator is not diagonal in the Mellin basis. In the logarithmic variable \(y=\log(k/\rho)\), \(R=i\partial_y\) and
\(P=\rho e^y\). Hence, at the level of generalized Mellin kernels
\begin{equation}
    \rho\, e^y \frac{e^{-i\omega y}}{\sqrt{2\pi}}
=
\rho \frac{e^{-i(\omega+i)y}}{\sqrt{2\pi}} \,.
\label{mellinkern1}
\end{equation}
Equivalently, one may write formally
\begin{equation}
    P\,|\omega\rangle_{\rm M} = \rho\,|\omega+i\rangle_{\rm M} .
\end{equation}

This equation is only a formal shorthand: \(|\omega+i\rangle_{\mathrm M}\) is not a Hilbert-space vector or an additional spectral eigenstate of the self-adjoint operator (R), but denotes the analytic continuation of the Mellin kernel in \eqref{mellinkern1}. More precisely, for Mellin wave packets whose transforms admit the relevant analytic continuation
\begin{equation}
    (P\psi)_{\rm M}(\omega)=\rho\,\psi_{\rm M}(\omega-i).
    \label{eq:complex omega}
\end{equation}
Thus the imaginary shift is a compact way of encoding the fact that null translations mix dilation eigenmodes. The analytic continuation in \ref{eq:complex omega} is consistent with the \textit{strip of convergence} of the Mellin transform, namely the values of complex $\omega$ for which the Mellin integral converges (see Appendix B). This Mellin description is the intrinsic dilation spectral representation of the affine particle. To connect it with the accelerated observer, however, one must also impose the geometric restriction to a Rindler half-line, where dilations remain unitary but translations no longer act as a unitary group.

\subsection{Restriction to the Rindler half-line}
\label{subsec:wedge-rindler-modes}

The affine representation above lives naturally on the full light-ray momentum orbit, whereas the Rindler observer has access to a single half-line in position space. In the right wedge \(v>0\), let us introduce the logarithmic Rindler null coordinate
\begin{equation}
\xi=\log(\alpha v),
\label{eq:xi-revised}
\end{equation}
where \(\alpha\) is the inverse length used in the definition of the Rindler coordinates.\\

The affine translation subgroup does not preserve the half-line \(v>0\): a finite translation \(v\mapsto v+\gamma\) can move points across the horizon. The dilation subgroup, however, does preserve \(v>0\). On
\[
\mathcal H_\xi=L^2(\mathbb R,d\xi),
\]
dilations act as translations of \(\xi\):
\begin{equation}
(\mathsf S_\xi(\lambda)F)(\xi)
=
F(\xi-\lambda),
\label{eq:dilation-xi-action}
\end{equation}
with generator
\begin{equation}
R_\xi=-i\frac{d}{d\xi}.
\label{eq:R-xi-revised}
\end{equation}
Thus the \(R_\xi\)-eigenfunctions are
\begin{equation}
\langle \xi|\omega\rangle_{\xi,R}
=
\frac{1}{\sqrt{2\pi}}e^{i\omega\xi},
\qquad
R_\xi|\omega\rangle_{\xi,R}
=
\omega|\omega\rangle_{\xi,R},
\label{eq:xi-R-eigenfunctions}
\end{equation}
and
\begin{equation}
{}_{\xi,R}\langle\omega|\omega'\rangle_{\xi,R}
=
\delta(\omega-\omega').
\label{eq:xi-orth-revised}
\end{equation}
The positive-frequency Rindler modes used in the field expansion are instead
\[
e^{-i\omega\xi}=v^{-i\omega},
\qquad
\omega>0.
\]
With the convention \eqref{eq:R-xi-revised}, these diagonalize the positive
Rindler Hamiltonian
\begin{equation}
H_{\rm R}:=-R_\xi=i\frac{d}{d\xi},
\qquad
H_{\rm R}e^{-i\omega\xi}=\omega e^{-i\omega\xi}.
\label{eq:Rindler-H}
\end{equation}
Equivalently,
\[
|\omega\rangle_{\xi,+}
=
|-\omega\rangle_{\xi,R},
\qquad
\langle\xi|\omega\rangle_{\xi,+}
=
\frac{1}{\sqrt{2\pi}}e^{-i\omega\xi}.
\label{eq:positive-rindler-mode}
\]
The translation generator on the half-line would be realized as
\begin{equation}
P_\xi=i\frac{d}{dv}
=
i\alpha e^{-\xi}\frac{d}{d\xi},
\label{eq:P-xi}
\end{equation}
for the \(e^{-ikv}\) convention. This operator is not self-adjoint on \(L^2(\mathbb R,d\xi)\), and the corresponding finite translations do not define a unitary group on the right wedge. Thus the wedge Hilbert space {\it carries only a unitary representation of the dilation subgroup, not of the full affine group.} This is analogous to the case of momentum in an infinite potential well in quantum mechanics.\\

To summarize, the affine one-particle space admits the Mellin eigenfunctions

\begin{equation}
\langle k|\omega\rangle_{\mathrm M}
=\frac{1}{\sqrt{2\pi}}k^{-i\omega},
\qquad
R|\omega\rangle_{\mathrm M}
=\omega|\omega\rangle_{\mathrm M},
\end{equation}
which are delta-normalized with respect to the affine inner product $\langle\cdot,\cdot\rangle_{\mathrm{aff}}$. On the Rindler half-line, the dilation generator \(R_\xi=-i\partial_\xi\) has instead the generalized eigenfunctions

\begin{equation}
\langle\xi|\omega\rangle_{\xi,R}
=\frac{1}{\sqrt{2\pi}}e^{i\omega\xi},
\qquad
R_\xi|\omega\rangle_{\xi,R}
=
\omega|\omega\rangle_{\xi,R},
\end{equation}

which are delta-normalized in \(\mathcal H_\xi=L^2(\mathbb R,d\xi)\).
Thus the two families provide spectral descriptions of the dilation generator in two different Hilbert-space realizations, with the same spectral parameter \(\omega\). The formal identification of their spectral coefficients is unitary, but it does not describe the restriction of Minkowski mode profiles to the Rindler half-line. The latter comparison contains a non-trivial spectral multiplier and is therefore not a unitary change of basis. The next section computes this comparison map explicitly.

\section{Mellin-Rindler overlap and the spectral comparison operator}
\label{sec:mellin-rindler-overlap}

Having identified the Mellin and Rindler bases as two dilation-diagonal descriptions of the same light-ray structure, we now proceed with their spectral comparison. This is the step at which the analytic distinction between
Minkowski and Rindler positive frequency becomes clear. For simplicity we set the reference scales \(\alpha=\rho=1\). Restoring them only multiplies the following kernels by phases depending on \(\omega\).

In the right wedge, \(v>0\), $v=e^\xi$ with $\xi\in\mathbb{R}$ Minkowski plane waves are represented in the \(\xi\)-basis by 
\begin{equation}
\langle \xi|k\rangle
= \frac{1}{\sqrt{2\pi}}\,
e^{-ik e^\xi}.
\end{equation}
Projecting a Mellin eigenstate onto the \(\xi\)-representation gives
\begin{align}
\langle\xi|\omega\rangle_{\rm M}
&=
\frac{1}{\sqrt{2\pi}}
\int_0^\infty\frac{dk}{k}\,
k^{-i\omega}e^{-ik e^\xi}
\nonumber\\
&=
\frac{1}{\sqrt{2\pi}}e^{i\omega\xi}
\int_0^\infty dp\,
p^{-i\omega-1}e^{-ip}.
\label{eq:Mellin-xi-overlap-step}
\end{align}
With the convergence prescription \(e^{-ip}\to e^{-(\varepsilon+i)p}\),
\(\varepsilon>0\), one obtains
\begin{equation}
\int_0^\infty dp\,
p^{-i\omega-1}e^{-ip}
=
e^{-\pi\omega/2}\Gamma(-i\omega).
\label{eq:gamma-overlap-minus}
\end{equation}
Therefore we can write
\begin{equation}
\langle\xi|\omega\rangle_{\rm M}
=
m_-(\omega)\,
\langle\xi|\omega\rangle_{\xi,R},
\qquad
m_-(\omega)
=
e^{-\pi\omega/2}\Gamma(-i\omega),
\label{eq:m-minus}
\end{equation}

where the subscript $-$ in the factor $m_-(\omega)$ labels the sign of the Minkowski plane-wave phase and should not be confused with the sign of the dilation frequency. Equivalently, for the positive-frequency Rindler modes \(e^{-i\omega\xi}\),
\(\omega>0\), one uses \(-\omega\) in \eqref{eq:m-minus}:
\begin{equation}
\langle\xi|-\omega\rangle_{\rm M}
=
m_-(-\omega)\,
\langle\xi|\omega\rangle_{\xi,+},
\qquad
m_-(-\omega)
=
e^{+\pi\omega/2}\Gamma(i\omega).
\label{eq:positive-rindler-overlap}
\end{equation}
The factor $m_-(\omega)$ is the same Gamma function-dependent factor that appears when a Minkowski plane wave is written as a Mellin superposition of Rindler modes:
\begin{equation}
e^{-ikv}
=
\frac{1}{2\pi}
\int_{-\infty}^{+\infty}d\omega\,
\Gamma(i\omega)e^{\pi\omega/2}
k^{-i\omega}v^{-i\omega},
\qquad k>0.
\label{eq:plane-wave-mellin-minus}
\end{equation}
Similarly we can write
\begin{equation}
e^{+ikv}
=
\frac{1}{2\pi}
\int_{-\infty}^{+\infty}d\omega\,
\Gamma(i\omega)e^{-\pi\omega/2}
k^{-i\omega}v^{-i\omega}.
\label{eq:plane-wave-mellin-plus}
\end{equation}

Substituting \eqref{eq:plane-wave-mellin-minus} and
\eqref{eq:plane-wave-mellin-plus} into the right-wedge field expansion
\eqref{psi +}, and comparing with \eqref{Rind pla exp}, gives the formal
Bogoliubov relation
\begin{equation}
B(\omega)
=
\frac{\sqrt{\omega}}{2\pi}\Gamma(i\omega)
\int_0^\infty\frac{dk}{\sqrt{k}}\,
k^{-i\omega}
\left[
e^{\pi\omega/2}a(-k)
+
e^{-\pi\omega/2}a^\dagger(-k)
\right],
\qquad
\omega>0.
\label{eq:B-in-terms-of-a-revised}
\end{equation}
The appearance of both \(a(-k)\) and \(a^\dagger(-k)\) reflects the usual Minkowski--Rindler mixing of positive and negative frequencies. In the present language, the reason is simple: Minkowski modes diagonalize translations along the null ray, while Rindler modes diagonalize dilations of that ray.\\

Returning to the relation \eqref{eq:m-minus} we can see that for a Mellin wave packet
\begin{equation}
|\psi\rangle_{\rm M}
=
\int d\omega\,\psi(\omega)|\omega\rangle_{\rm M},
\end{equation}
the corresponding right-wedge profile is
\begin{equation}
(I_-\psi)(\xi)
=
\int d\omega\,
m_-(\omega)\psi(\omega)
\frac{e^{i\omega\xi}}{\sqrt{2\pi}}.
\label{eq:I-minus-wavepacket}
\end{equation}

\begin{figure}[t]
    \centering
    \includegraphics[width=\textwidth]{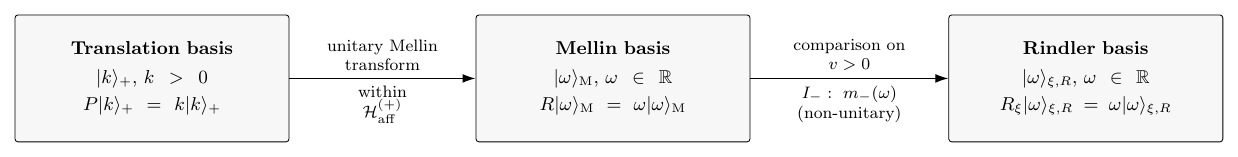}
    \caption{Schematic relation between the translation, Mellin and Rindler
    descriptions. The Mellin transform is a unitary change of spectral
    representation within the affine one-particle space. Comparison with the
    Rindler modes on a single half-line instead introduces the non-unitary
    multiplier $m_-(\omega)$, whose opposite-frequency weights contain the
    thermal factor.}
    \label{fig:affine-bases}
\end{figure}

This equation extends the relation between generalized dilation eigenfunctions to Mellin wave packets. To make precise when this expression defines a right-wedge profile, we regard \(I_-\) as a map
\begin{equation}
I_-:D(I_-)\subset L^2(\mathbb R,d\omega)
\longrightarrow L^2(\mathbb R,d\xi),
\end{equation}
where
\begin{equation}
D(I_-)=\{
\psi\in L^2(\mathbb R,d\omega):
m_-(\omega)\psi(\omega)\in L^2(\mathbb R,d\omega)\}.
\end{equation}
This domain is dense: wave packets which vanish in a neighbourhood of \(\omega=0\) belong to \(D(I_-)\) and can approximate any square-integrable wave packet. Since \(m_-(\omega)\) has a pole at \(\omega=0\), the map is unbounded. Equation \eqref{eq:I-minus-wavepacket} also makes precise the sense in which \(I_-\) is an intertwiner. Under a dilation by \(\lambda\), the Mellin coefficients acquire the phase \(e^{-i\lambda\omega}\), and one finds
\begin{equation}
e^{-i\lambda R_\xi}I_-\psi
= I_-\left(e^{-i\lambda\omega}\psi\right),
\qquad
\psi\in D(I_-).
\end{equation}
Thus \(I_-\) intertwines the two realizations of the dilation subgroup. It does not intertwine the full affine group, since translations do not act unitarily on the Rindler half-line. The relations between the three spectral descriptions discussed above are
summarized in Fig.~\ref{fig:affine-bases}. For \(\psi\in D(I_-)\), Eq.~(55) gives
\begin{equation}
|I_-\psi|_{L^2(d\xi)}^2
=\int_{\mathbb R}d\omega
|m_-(\omega)|^2|\psi(\omega)|^2.
\label{eq:pullback-norm}
\end{equation}
This relation does not introduce an independent inner product on the Mellin space: it is simply the Rindler norm evaluated on the image of the wave packet under \(I_-\). Since \(|m_-(\omega)|\) is not a pure phase, this norm is not preserved and the comparison map is not unitary.

This observation is closely related to the distinction between inertial and Rindler particles. Minkowski particles are defined by positive frequency with respect to inertial time, whereas Rindler particles are defined by positive frequency with respect to boost time inside a wedge. Restricting a Minkowski positive-frequency mode to a
single wedge is therefore not a unitary change of basis within the positive Rindler-frequency sector. The analytic structure of the Minkowski mode fixes a definite relative weight between the two signs of the dilation frequency. For \(\omega>0\),
\begin{equation}
|m_-(\omega)|^2
=
e^{-\pi\omega}|\Gamma(i\omega)|^2,
\qquad
|m_-(-\omega)|^2
=
e^{\pi\omega}|\Gamma(i\omega)|^2,
\end{equation}
and hence
\begin{equation}
\frac{|m_-(\omega)|^2}{|m_-(-\omega)|^2}
=
e^{-2\pi\omega}.
\label{eq:thermal-ratio}
\end{equation}

This is the characteristic Boltzmann factor associated with the Rindler temperature \(T=1/(2\pi)\), in units where the acceleration scale has been set to one. In this sense the multiplier \(m_-(\omega)\) contains, already at the level of the one-particle mode functions, the analytic imbalance between positive and negative Rindler frequencies which underlies the Unruh effect.
In terms of probabilities, we have that $P_{\rm M}(\omega) = |\bra{\xi}\ket{\omega}_{\rm M}|^2 = |m_-(\omega)|^2 \, |\bra{\xi}\ket{\omega}_{\xi, R}|^2 = \frac{2\pi}{\omega} \frac{1}{e^{2 \pi \omega}-1} P_R(\omega)$, so there is a mismatch between $P_{\rm M}(\omega)$ and $P_R(\omega)$: the Rindler observer needs to properly rescale probabilities by an $\omega$-weighting factor related to Bose-Einstein distribution. The rescaling of probabilities acts just as if there was a variable number of $\omega$-quanta, weighted by the thermal distribution factor $|m_-(\omega)|^2$. Indeed, $|{}_{\xi, R}\bra{\omega} \ket{\omega'}_{\rm M}|^2 = \frac{2\pi}{\omega} \frac{1}{e^{2 \pi \omega}-1} \delta(\omega-\omega')$ implies that one of these states, although pure relative to its corresponding observer (mathematically, with respect to the Hilbert space it is defined in), exhibits Planckian spectrum with respect to the other's  modes. The factor proportional to $1/\omega$ represents the density of states, or the so-called phase-space measure.

The point \(\omega=0\) is singular because of the pole of the Gamma function. This is the usual zero-mode subtlety of the massless scalar field and should be treated separately. As a matter of fact, the massless two-point function in $d$ spacetime dimensions is given by $\langle \phi(x) \phi(y) \rangle = \frac{\Gamma(\frac{d}{2}-1)}{4 \pi^{d/2} ||x-y||^{d-2}}$ which leads to an IR divergence for $d=2$ due to the presence of $\Gamma(0)$. This poses no particular issue, since observables in the theory are built from (shift-invariant) derivatives of the field which are well-behaved in the vacuum. The algebraic approach, which we will review in the following section, holds the advantage of describing the theory directly in terms of the physical observables. Away from this point, the conclusion is that the Mellin and Rindler dilation bases are diagonal with respect to the same dilation parameter, but they are not related by a unitary change of delta-normalized basis.

The calculation in this section shows how the Unruh factor arises at the one-particle level from the analytic comparison between translation-adapted and dilation-adapted modes. To understand why the same dilation flow defines a thermal equilibrium condition for the wedge algebra, we now turn to modular theory.

\section{Modular Hamiltonian, analyticity and KMS condition}


The one-particle calculation above identifies the dilation flow as the spectral flow responsible for the Rindler thermal factor. Modular theory explains the operator-algebraic meaning of this statement. For a standard half-line algebra and a suitable cyclic and separating state, the modular flow is implemented by dilations, and the KMS condition expresses the thermal
character of the state restricted to the wedge. In this section we connect the affine representation-theoretic picture to this modular description. This connection constitutes the main extension of the representation-theoretic analysis sketched in \cite{Arzano:2025aeo}. The purpose is to show how the dilation generator responsible for the thermal spectral factor at the one-particle level is embedded in the established modular structure of the horizon half-line algebra.

We shall begin with a heuristic argument for the link between analytic properties of certain operators and the corresponding thermal behavior. Consider the operator $f_R(\lambda)=e^{i \lambda R}$: if we think of extending $f_R(\lambda) = \int d\omega f_{\omega}(\lambda) \ket{\omega} \bra{\omega}$ to complex $\lambda$,

\begin{equation}
    e^{i \left( \mathfrak{Re}(\lambda) + i \mathfrak{Im}(\lambda) \right) (\omega-i \sigma)} = e^{\omega \left(- \mathfrak{Im}(\lambda) + i\mathfrak{Re}(\lambda) \right)} e^{\sigma \left( \mathfrak{Re}(\lambda) + i \mathfrak{Im}(\lambda) \right)}
    \label{eq:complex a}
\end{equation}

imposes some restrictions on the possible values of $\mathfrak{Im}(\lambda)$. In particular, in order for the integral \ref{eq:complex a} to converge, we should impose $\mathfrak{Im}(\lambda) > 0$ as we integrate for arbitrarily high values of $\omega > 0$ in the spectral decomposition, while $\mathfrak{Im}(\lambda) < 0$ for $\omega < 0$. If an observer localized in a certain region $\mathcal{O}$ has $\pm \left. R \right|_{\mathcal{O}}$ as their Hamiltonian, the positivity condition would require $\pm \omega > 0$, so she may analytically continue to $\pm \mathfrak{Im}(\lambda) > 0$. This range of values of $\omega$ for which we can perform the analytic continuation to complex $\lambda$ corresponds precisely to a half-line, such as $v>0$.

Additionally, physical quantities will generally be computed from $f_R(\lambda)$. If we wanted to study the general analytic properties of such observables for complex values of both $\lambda$ and $\omega$, we may analytically continue $R$: if we analytically continued $\omega$ and integrated over arbitrarily high values of $\sigma$ as well, corresponding to $f_{R - iR}(\lambda)$, we would have a restriction on $\mathfrak{Re}(\lambda)$, which conflicts with the fact that $\mathfrak{Re}(\lambda) \in \mathbb{R}$ is the dilation parameter. As a result, $\sigma$ must belong at most to an interval $(a, b)$, as compatible with the strip of convergence of the Mellin integral. This also provides a heuristic insight into the convergence domain of the Mellin transform, from purely operator-algebraic considerations.

The analytic constraint on $f_R(\lambda)$ is analogous to the case of Rindler wedges, where the Killing vector field $\xi_K$ associated to boosts' generator $K$ is timelike only in the wedges and flips direction going from the right wedge $\mathcal{R}$ to the left one $\mathcal{L}$: $\xi_K = x \partial_t + t \partial_x \cong (x, t, 0, 0)$ implies, in the $(-,+,+,+)$ signature,

\begin{equation}
    \begin{cases}
        &\xi_K^0 > 0 \\
        &||\xi_K||^2 < 0
    \end{cases} \, \, \, \text{in $\mathcal{R}$,} \qquad 
    \begin{cases}
        &\xi_K^0 < 0 \\
        &||\xi_K||^2 < 0
    \end{cases} \, \, \, \text{in $\mathcal{L}$.}
\end{equation}

In that case, the associated charge evaluated on the $t=0$ hypersurface is

\begin{equation}
    K = \int_{\mathbb{R}^3} dx \, dy \, dz \, T_{00} \, x = \int_{\mathbb{R}^2} dy \, dz \int_0^{\infty} dx \, T_{00} \, x + \int_{\mathbb{R}^2} dy \, dz \int_{-\infty}^{0} dx \, T_{00} \, x = K_\mathcal{R} - K_\mathcal{L}
\end{equation}

with $K_{\mathcal{R}, \mathcal{L}} > 0$. Upon quantization\footnote{Strictly speaking, this does not hold in QFT because of the unsplitting of $K$ into $K_\mathcal{R} - K_\mathcal{L}$, due to wedge algebras being type $\mathrm{III}$. The mathematical reason is that the modular automorphism group cannot be inner in a type $\rm III$ algebra. A physical explanation is that $K_\mathcal{R} \ket{\Omega}$ does not belong to the Hilbert space, as $\bra{\Omega} K_\mathcal{R}^2 \ket{\Omega} = \infty$; when integrated over all space, the divergent contributions cancel out since $K \ket{\Omega} = 0$. As a result, although $K$ is a well-defined operator, the single pieces $K_{\mathcal{R}, \mathcal{L}}$ are not. Notwithstanding, the argument about the analyticity still applies.}, one expects $K_{\mathcal{R}, \mathcal{L}}$ to have positive spectrum. 
The information about the region restriction, i.e. the observer experiencing thermality, comes from the analyticity condition on $e^{i \eta K}$ for complex $\eta$. For the purpose of this work, the discussion was kept rather informal, but the argument can be made rigorous with the theory of operator algebras. In fact, the analyticity properties relate to the wedge restriction through Bisognano-Wichmann theorem. Specifically, to keep the modular flow being an automorphism in agreement with Tomita-Takesaki theorem, one needs to restrict the algebra of field observables to a wedge, since the modular flow is ergodic in such region.

Consider a $1+1$ real, massless scalar field theory. Given a (von Neumann) algebra of operators, the Tomita operator can be defined by the relation

\begin{equation}
    S_{\Omega} A \ket{\Omega} = A^{\dagger} \ket{\Omega}
\end{equation}

for any element $A$ in the algebra. In the case of QFT, fields are operator-valued distributions. To obtain properly-defined operators, we smear the field with a complex test function $f(t,x)$, with compact support in a certain region of Minkowski spacetime, obtaining the generally unbounded operator $\phi[f] \equiv \int dt dx \phi(t,x) f(t,x)$. $A$ can be thought of as a bounded function of the smeared field operator $\phi[f]$, such as a Weyl-like exponentiation. Applying known lemmas from calculus of variation, we end up with the equation for the field operator

\begin{equation}
    S_{\Omega} \phi(t,x) \ket{\Omega} = \phi(t,x) \ket{\Omega}.
\end{equation}

Notice that $S_{\Omega}$ cannot be the identity as it is antilinear: the operator must send $\phi[f] \ket{\Omega}$ to $\phi[f^*] \ket{\Omega}$. Since in lightcone coordinates we can split $\phi(t,x) = \phi(u, v) = \chi(u) + \psi(v)$, the previous relation becomes

\begin{equation}
    S_{\Omega} \psi(v) \ket{\Omega} = \psi(v) \ket{\Omega}
\end{equation}

and analogously for $\chi(u)$. The modular operator is defined in terms of the Tomita operator by the relation $\Delta_{\Omega}=S_{\Omega}^{\dagger} S_{\Omega}$. Such operator can be computed explicitly if we notice that

\begin{equation}
    \Theta_v e^{-\pi R} \psi(v) \ket{\Omega} = \Theta_v \psi(-v) \ket{\Omega} = \psi(v) \ket{\Omega} = S_{\Omega} \psi(v) \ket{\Omega}
    \label{eq:Modular operator}
\end{equation}
so that we can write
\begin{equation}
    \Delta_{\Omega} = e^{-2 \pi R}
\end{equation}
and $J_{\Omega} = \Theta_v$, where $\Theta_v$ denotes a reflection on the $v$-line. It can be checked that $\Theta_v e^{-\pi R}$ does indeed take a generic $A \ket{\Omega}$ to $A^{\dagger} \ket{\Omega}$. This happens because boosts, which are the modular flow of a Rindler wedge, become dilations on the lightcone. 

Notice that equation \ref{eq:Modular operator} explicitly assumes the analytic continuation of $e^{i \lambda R}$ to $\lambda = i \pi$. The analytic domain of $f_R(\lambda)$ is essential in the discussion and it is what forces to restrict ourselves to the half-line. Indeed, the spectrum of $R$ is continuous and unbounded as manifest from representation theory; in order for the operator $f_R(\lambda)$ to make sense for complex $\lambda$, we need to restrict to a region where $R$ has bounded spectrum at least from one end.
On \(L^2(\mathbb R,d\xi)\) the generator of translations in \(\xi\) has spectrum \(\mathbb R\). The half-line restriction selects the algebra on which dilations act geometrically as automorphisms, and the KMS condition is a statement about the analytic continuation of correlation
functions in a strip. The positive-frequency Rindler sector then corresponds to a choice of spectral splitting for the Rindler Hamiltonian, which allows us to write \ref{eq:Modular operator}.

Going back to equation \ref{eq:Modular operator}, we can now compute the action of $S_{\Omega}$ on the representation explicitly. $\Delta_{\Omega} = e^{-2 \pi R}$ is trivially diagonal on $\ket{\omega}_{\rm M}$, hence its action is known. To check the role of $J_{\Omega}$ in the representation, notice that the field $\psi(v)$ can be decomposed in terms of the unitary irreducible representations $L^2(\mathbb{R}_{\pm}, \frac{dk}{k})$ of the $ax+b$ group as

\begin{equation}
    \begin{split}
        \psi(v) &= \bra{v} \ket{\psi} = \frac{1}{\sqrt{2 \pi}} \int_0^{\infty} \frac{dk}{k} e^{ikv} {}_{+}\bra{k}\ket{\psi} + \frac{1}{\sqrt{2 \pi}} \int_{-\infty}^{0} \frac{dk}{|k|} e^{ikv} {}_{-}\bra{k}\ket{\psi} \\
        &= \frac{1}{\sqrt{2 \pi}} \int_0^{\infty} \frac{dk}{k} \left( e^{ikv} a(k) + e^{-ikv} a^{\dagger}(k) \right)
    \end{split}
\end{equation}

with $a(k) = {}_{+}\bra{k}\ket{\psi}$. Since the field is real, ${}_{+}\bra{k}\ket{\psi} = {}_{-}\bra{-k}\ket{\psi}^*$ and we can put ourselves in one irreducible representation, say the $\mathbb{R}_+$ one:

\begin{equation}
    \begin{split}
        S_{\Omega} \psi(v) \ket{\Omega} &= \Theta_v e^{-\pi R} \frac{1}{\sqrt{2 \pi}} \int_0^{\infty} \frac{dk}{k} e^{-ikv} a^{\dagger}(k) \ket{\Omega} \\
        &= \frac{1}{\sqrt{2 \pi}} \int_0^{\infty} \frac{dk}{k} \Theta_v e^{-\pi R} e^{-ikv} \sqrt{k} \ket{k}_+ \\
        &= \frac{1}{\sqrt{2 \pi}} \int_0^{\infty} \frac{dk}{\sqrt{k}} e^{-ikv} \Theta_v \ket{-k}_+
    \end{split}
\end{equation}

which must be equal to $\frac{1}{\sqrt{2 \pi}} \int_0^{\infty} \frac{dk}{\sqrt{k}} e^{-ikv} \ket{k}_+$. Hence the action of the modular conjugation $J_{\Omega} = \Theta_v$ on one-particle states is given by

\begin{equation}
    \Theta_v \ket{k}_{-}^* = \Theta_v \ket{-k}_+ = \ket{k}_+
    \label{J on 1 particle}
\end{equation}

which is a representation intertwiner between $L^2(\mathbb{R}^{\pm}, \frac{dk}{k})$. The $*$ is to be interpreted as the dual action $\ket{\varphi}^* = \bra{\varphi}$\footnote{If $\ket{b} = \ket{a}^*$, their components are complex conjugates to each other: $\bra{b}\ket{e_n}=\bra{e_n}\ket{a} \implies a_n = b_n^*$.}; as a result, $J_{\Omega} = \Theta_v$ acts on states as a complex conjugation, analogously to the action of time reversal in quantum mechanics. Naturally, $S_{\Omega}$ cannot act trivially on $\ket{k}$ since $S_{\Omega}$ cannot be the identity.

The KMS condition for the field $\psi(v)$ restricted on $v>0$ reads

\begin{equation}
    \bra{\Omega} e^{i \lambda R} \psi(v) e^{-i \lambda R} \psi(v') \ket{\Omega} = \bra{\Omega} \psi(v') e^{i(\lambda +i \beta)R} \psi(v) e^{-i(\lambda +i \beta)R} \ket{\Omega}
\end{equation}

that is

\begin{equation}
    \bra{\Omega} \psi(e^{\lambda} v) \psi(v') \ket{\Omega} = \bra{\Omega} \psi(v') \psi(e^{i \beta} e^{\lambda} v) \ket{\Omega}.
\end{equation}

As a consequence,

\begin{equation}
    \bra{v}\ket{v'} = \bra{v'} \ket{e^{i \beta} v}
\end{equation}

telling us that, at one-particle level, the representations of the $ax+b$ group must satisfy a certain constraint about the analytic form of the propagator.

The reason why the modular properties of the field algebra are related to the one-particle structure of the theory is not a coincidence. It is a general feature of quantized fields, mathematically due to a correspondence between the one-particle Hilbert space $\mathcal{H}$ and the Fock space\footnote{Bosonic, in this case.} $\Gamma(\mathcal{H}) \equiv \bigoplus_{n=0}^{\infty} \mathrm{Sym} (\mathcal{H}^{\otimes n})$ \cite{Longo:inbook} \cite{Longo:lecture_notes_1} \cite{Longo:lecture_notes_2}. Summarizing the result, one can develop modular theory for a \textit{standard subspace} $H \subset \mathcal{H}$, meaning that $H$ is a real subspace of $\mathcal{H}$ such that $\overline{H + i H} = \mathcal{H}$ and $H \cap i H = \{ 0 \}$. It can be proven that a suitable Tomita operator $S_H$ associated to the inclusion $H \subset \mathcal{H}$ can be defined and the corresponding operators $\Gamma(\Delta_H)$ and $\Gamma(J_H)$ respectively coincide with the operators $\Delta_{\mathcal{A}(H)}$ and $J_{\mathcal{A}(H)}$ obtained via modular theory for the von Neumann algebra $\mathcal{A}(H) = \{ W(\xi) : \xi \in H \}''$ which acts on $\Gamma(\mathcal{H})$. Here $H$ is to be interpreted\footnote{The reason why this functional space is included in $\mathcal{H}$ is because by acting on the vacuum with $\phi[\xi]$ we get one-particle states. The standard subspace properties are trivially satisfied, since $iH$ contains imaginary-valued functions and $H+iH$ is dense in $\mathcal{H}$ due to the Reeh-Schlieder theorem.} as the space of real test functions with compact support in a region $\mathcal{O}$, while $W$ denotes the associated Weyl operator; consequently $\mathcal{A}(H) \equiv \mathcal{A}(\mathcal{O})$ in algebraic QFT. In our case, $J_{\mathcal{A}(H)}$ would be $J_{\Omega} = \Theta_v$ since we derived it directly from the field algebra, but one may construct a corresponding $J_H$ such that $\Theta_v = \Gamma(J_H)$. To clarify, equation \ref{J on 1 particle} provides the action of $J_{\mathcal{A}(H)}$ on the one-particle Hilbert space, which coincides with $J_{H}$ because $\left. J_{\mathcal{A}(H)} \right|_{\mathcal{H}} = \left. \Gamma(J_H) \right|_{\mathcal{H}} = J_H$: when restricted to the one-particle sector, their actions coincide; their difference lies however in the domains they act on (one-particle sector versus many-particles). As a result, $J_H \ket{k}_{-}^* = \ket{k}_{+}$. A physical explanation to the fact that modular information is encoded in the one-particle structure is that a quantum field is ultimately a model of many identical particles.

The above considerations, besides linking the modular treatment of the one-particle structure to the explicit calculations performed above with the Mellin transform, also suggest that the information about the Poincaré structure to define the vacuum $\ket{\Omega}$ is redundant when we restrict ourselves to the line: we can extract thermality from the representations of the $ax+b$ group alone. As a matter of fact, given an algebra $\mathcal{A}_+$ localized on the positive half of the $v$-line, we can define a suitable vector $\ket{\Omega}$ as the GNS representative of a $R$- and $P$-invariant state on $\mathcal{A}_+$, without specifying any underlying Poincaré structure. Assuming such state exists\footnote{And it does when we see the theory on the line as arising from a $1+1$ theory restricted on the lightray, since it is the Minkowski vacuum. What we are trying to remark here is that the reference to the underlying $2$d theory is not necessary, since we can construct such desired state algebraically.}, the KMS condition follows from modular theory. The fact that the Poincaré group is not necessary strongly suggests that a similar result can be obtained in any spacetime characterized by the $ax+b$ group near a bifurcate Killing horizon.

A way to construct such state is to proceed backwards with respect to what we did above. Starting from a representation $\psi(v)$ built from $L^2(\mathbb{R}_{\pm}, \frac{dk}{k})$, consider another copy of the original representation, i.e. define another field $\chi(u)$ in addition to $\psi(v)$, then combine them as $\phi(u,v) = \chi(u) + \psi(v)$ and finally define $\ket{\Omega}$ as a the $\phi$-vacuum. As trivial as it may seem, in this case we actually do not start from the existence of the vacuum, but rather derive it from the representation theory of $ax+b$.

An equivalent version for this observation can be provided by Borchers' theorem \cite{Borchers:1991xk, Summers:1995kp}, which can be used to perform the backwards construction outlined above in algebraic terms \cite{Longo:2010we, Bischoff:2013dha}. Consider a von Neumann algebra $\mathcal{M}$ acting on a Hilbert space $\mathcal{H}$ with cyclic and separating vector $\ket{\Omega}$; if there is a continuous one-parameter group of unitaries $U(\gamma)$ leaving $\ket{\Omega}$ invariant such that its generator is positive definite, then for all $t, \gamma \in \mathbb{R}$:

\begin{enumerate}
    \item if $U(\gamma_+) \mathcal{M} U^{\dagger}(\gamma_+) \subseteq \mathcal{M}$ for $\gamma_+ \geq 0$, then $\Delta_{\Omega}^{-it} U(\gamma) \Delta_{\Omega}^{it} = U(e^{-2 \pi t} \gamma)$ and $J_{\Omega} U(\gamma) J_{\Omega} = U(- \gamma)$;
    \item if $U(\gamma_-) \mathcal{M} U^{\dagger}(\gamma_-) \subseteq \mathcal{M}$ for $\gamma_- \leq 0$, then $\Delta_{\Omega}^{-it} U(\gamma) \Delta_{\Omega}^{it} = U(e^{2 \pi t} \gamma)$ and $J_{\Omega} U(\gamma) J_{\Omega} = U(- \gamma)$.
\end{enumerate}

\begin{figure}[h!]
    \centering
    \includegraphics[width=0.675\textwidth]{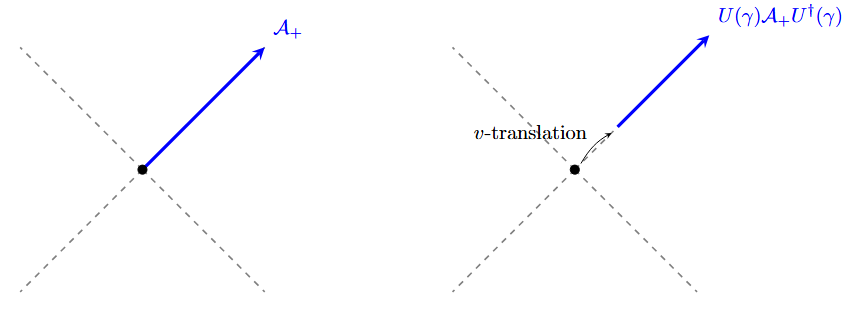}
    \caption{Visualization of the algebra $\mathcal{A}_+$ and its $v$-translated $U(\gamma) \mathcal{A}_+ U^{\dagger}(\gamma)$ for positive $\gamma$. Notice that $U(\gamma) \mathcal{A}_+ U^{\dagger}(\gamma) \not\subseteq \mathcal{A}_+$ for $\gamma < 0$, since points are moved across the horizon (in this sense, the origin).}
    \label{figure}
\end{figure}

As a result, the modular operator $\Delta_{\Omega}^{it}$ acts as $\gamma$-dilations. Identifying $\mathcal{M} = \mathcal{A}_+$ as the algebra restricted on the positive $v$ half-line, $U(\gamma)$ implements the null $v$-translations, as they geometrically yield a shifted half-line which is properly included in the original one, satisfying the first of Borchers' conditions above (see figure \ref{figure}). Automatically, we deduce $\Delta_{\Omega} = e^{-2 \pi R}$, $J_{\Omega} = \Theta_v$. An analogous description can be done for the negative $v$ half-line, where now the second condition is satisfied, and similarly for $u$.

The modular analysis therefore completes the affine interpretation of the Rindler horizon, also providing an insight between the Mellin transform and modular theory. The Mellin calculation shows how the thermal factor appears in the comparison of one-particle mode bases, while modular theory identifies the same dilation flow as the automorphism group with respect to which the restricted vacuum is a KMS state. The recent developments in the modular theory of standard subspaces also confirm the idea that one-particle modes contain all the sufficient thermal information, as explicitly manifest via the Mellin computation. The common element in both descriptions is the affine structure of the light ray: translations select the inertial positive-frequency splitting, dilations select the Rindler splitting, and the half-line algebra turns the dilation flow into modular dynamics.


\section{Conclusion and outlook}

We have revisited the Unruh effect from the viewpoint of the affine symmetry of a horizon light ray.  For a massless scalar field in two spacetime dimensions, the light-cone decomposition isolates independent chiral sectors, each naturally associated with a null coordinate. On such a null ray, Minkowski modes are adapted to translations, whereas Rindler modes are adapted to dilations. These two transformations generate the one-dimensional affine group, so the distinction between inertial and accelerated particles can be
understood as the distinction between two spectral decompositions associated with two non-commuting generators of the same symmetry.

The one-particle realization of this statement is provided by the positive continuous irreducible representation of the affine group on the positive light-cone momentum orbit. A central point of the analysis is that the restriction to a Rindler wedge does not define a unitary change of positive-frequency basis. Translations along the full light ray do not preserve the Rindler half-line, while dilations do. 

We have then related this one-particle picture to modular theory. The modular description shows that the dilation flow is not merely a convenient single-particle diagonalization: for the half-line algebra it is the modular flow with respect to the vacuum state, and the corresponding KMS condition expresses the thermal character of the restricted vacuum. The affine
calculation and the modular-theoretic statement therefore describe the same phenomenon from complementary perspectives. The former exhibits the analytic origin of the thermal factor at the level of modes; the latter identifies the operator-algebraic reason why the restricted state is thermal.

The explicit construction presented here applies to a massless scalar field in two-dimensional Minkowski spacetime. For a massive scalar, boosts and null translations still generate an affine subgroup and the Unruh temperature is unchanged, but the mass couples the left- and right-moving sectors and the Rindler modes are no longer pure dilation waves of a single null coordinate. The present construction is recovered most directly in the near-horizon regime, where the mass term becomes subleading. In higher dimensions, transverse momenta play the role of an effective mass, while the full mode structure depends on the geometry.

For a massless fermion, an analogous chiral affine description survives, with the appropriate spinorial scaling factors; the thermal occupation numbers are then of Fermi--Dirac rather than Bose--Einstein form. For a massive fermion the exact chiral reduction is again lost, although the same near-horizon reasoning applies. More generally, near a smooth non-extremal Killing horizon the normal two-dimensional geometry is locally Rindler, with the surface gravity fixing the normalization of the dilation flow and hence the temperature. An operator-algebraic extension nevertheless requires a suitable horizon or wedge algebra and a state for which the modular flow is geometric. The affine near-horizon structure alone therefore does not imply a KMS statement in an arbitrary curved or non-stationary spacetime.

Finally, it would be interesting to develop the modular part of the construction in a fully algebraic form for more general spacetimes, using half-sided modular inclusions and Borchers' theorem as the starting point rather than as a posteriori interpretation. In that formulation, the affine group would emerge directly from the modular data
of a half-line algebra and a translation-invariant cyclic and separating state. This could provide a minimal algebraic route to horizon thermality and clarify which ingredients of horizon thermality are genuinely local to the horizon algebra and which ones instead depend on the existence of a global Poincaré-covariant quantum field theory. We leave a detailed exploration of these directions to future work.

\section*{Data availability statement}

This study is entirely theoretical and does not involve the generation or analysis of empirical data. Therefore, no data are available to share.

\section*{Acknowledgments}
We acknowledge support from the INFN Iniziativa Specifica QUAGRAP and from the European COST Actions BridgeQG CA23130 and CaLISTA CA21109. P.P. also acknowledges support from the European COST Action RQI CA23115.

\bibliography{biblio}

\onecolumngrid

\appendix

\section{Modular theory and KMS condition}

It is instructive to introduce modular theory through quantum-mechanical examples.
Consider an entangled state in quantum mechanics, say for instance a vector

\begin{equation}
    \ket{\omega} = \cos{\left( \frac{\theta}{2} \right)} \ket{0}_A \ket{0}_B + \sin{\left( \frac{\theta}{2} \right)} e^{i \varphi} \ket{1}_A \ket{1}_B
\end{equation}

in a two-qubit system. This vector yields a mixed state when restricted to $A$'s degrees of freedom:

\begin{equation}
    \rho_A = \cos^2{\left( \frac{\theta}{2} \right)} \ket{0}_A \bra{0}_A + \sin^2{\left( \frac{\theta}{2} \right)} \ket{1}_A \bra{1}_A.
\end{equation}

We may wonder whether there exists a suitable Hamiltonian $H_{\text{mod}}$ such that the mixed state $\rho_A$ looks like a thermal equilibrium state with respect to the dynamics induced by $H_{\text{mod}}$. In this simple scenario, the answer is trivially yes: we may impose

\begin{equation}
    \rho_A = \frac{e^{-\beta H_{\text{mod}}}}{Z}
\end{equation}

and taking the logarithm we have

\begin{equation}
    \beta H_{\text{mod}} = - \log{(\rho_A)} + \ \text{constant} = - \begin{bmatrix}
    \log{\cos^2{\left( \frac{\theta}{2} \right)}}  & 0 \\
    0 & \log{\sin^2{\left( \frac{\theta}{2} \right)}} 
    \end{bmatrix} + \text{constant}.
\end{equation}

$H_{\text{mod}}$ is called \textit{modular Hamiltonian}. The logarithm exists\footnote{$\rho_A$ is positive semi-definite and self-adjoint, so its logarithm exists when it has no zero eigenvalue.} for $\theta \notin \mathbb{Z} \pi$, that is whenever the state $\ket{\omega}$ is not entangled. In fact, for such values of $\theta$ we would have a pure density matrix $\rho_A$, which may never look thermal, since a thermal state is intrinsically mixed and decoherent.

In higher dimensions, whenever a state is \textit{fully entangled} \cite{RevModPhys.90.045003}, i.e. it admits a Schmidt decomposition

\begin{equation}
    \ket{\omega} = \sum_{i} \lambda_i \ket{i}_A \ket{i}_B
    \label{eq:fully entangled}
\end{equation}

with $\lambda_i \neq 0$, $\forall i$, there is a modular Hamiltonian associated to $\rho_A$.
One may wonder if this feature of entangled states applies generally. For instance, is there a similar result for local observables in quantum field theory?

First of all, it is known that local observables differ drastically from the algebras characterizing ordinary quantum mechanics. Observables in non-relativistic quantum mechanics are described mathematically in terms of type $\mathrm{I}$ von Neumann algebras. A von Neumann algebra is an algebra $\mathcal{A}$ of bounded operators closed in the weak topology. An algebra is said to be type $\mathrm{I}$ if and only if it is isomorphic to the algebra of all bounded operators $\mathcal{B}(\mathcal{H})$ acting on a suitable Hilbert space. In quantum mechanics, where the tensor product axiom holds, we have for two systems $A$ and $B$ that the global Hilbert space decomposes as $\mathcal{H} = \mathcal{H}_A \otimes \mathcal{H}_B$. The algebra of global observables is $\mathcal{B}(\mathcal{H})$, hence it is a type $\mathrm{I}$ algebra; the same holds for the subsystems $A$ and $B$, since their observables are respectively $\mathcal{B}(\mathcal{H}_A) \otimes \mathds{1} \cong \mathcal{B}(\mathcal{H}_A)$ and $\mathds{1} \otimes \mathcal{B}(\mathcal{H}_B) \cong \mathcal{B}(\mathcal{H}_B)$, thus also type $\mathrm{I}$ algebras. The picture changes dramatically in quantum field theory.

Local field algebras form indeed a type $\mathrm{III}$ von Neumann algebra, reflecting the UV-divergent nature of QFT. Specifically, a feature of this type of operator algebra is the impossibility of associating density matrices to observers who have causal access to proper subregions of spacetime. This is due to the fact that the Hilbert space of a field theory does not split into a tensor product $\mathcal{H} \neq \mathcal{H}_A \otimes \mathcal{H}_B$, when $A$ and $B$ are associated to two complementary spacetime subregions, such as two wedges in Minkowski spacetime. In a sense, the tensor product axiom fails when we consider spacetime regions as \textit{subsystems}. 

The type $\mathrm{III}$ nature of local algebras is not just a formal mathematical feature, but hides deep physical consequences. An operational definition can be given in terms of the embezzling power of QFT, as found in recent developments \cite{Luijk:2024}. Essentially, the type classification of von Neumann algebras clarifies the entanglement structure of the theory.

As a result, the above argument about the existence of a modular Hamiltonian fails, at least in the methodology. Notwithstanding, we expect an analogous physical scenario to occur, since we know that observers characterized by a horizon measure a finite temperature.

To get around the obstruction arising from the non-factorization of the Hilbert space into a tensor product, we can move to an algebraic description. First, we should formulate the concept of \textit{entanglement} in algebraic terms. We define a vector $\ket{\omega}$ to be \textit{fully entangled} if and only if it is both \textit{cyclic} and \textit{separating} for the algebra of observables $\mathcal{A}$ considered. Cyclic means that any vector in the Hilbert space can be approximated by acting on the considered state with a suitable operator in the algebra: $\overline{\mathcal{A} \ket{\omega}} = \mathcal{H}$; separating means that if an operator annihilates the state, then it must be identically zero: $A \ket{\omega} = 0 \implies A = 0$. The vacuum satisfies this property due to the Reeh-Schlieder theorem. In the case of a type $I$ algebra, this definition is equivalent to \ref{eq:fully entangled}.

Secondly, we need to generalize the notion of \textit{thermal state} when we lack density matrices at our disposal. A way to bypass the use of density matrices is to define thermal states in terms of the properties of correlation functions. In the type-$\mathrm{I}$ case, a characterization of thermal states is the following behaviour of the correlation functions, called KMS condition:

\begin{equation}
    \begin{split}
        \langle A(t) B \rangle_{\beta} &= \Tr{\rho_{\beta} e^{iHt} A e^{-iHt} B} = \frac{1}{Z} \Tr{e^{-\beta H} e^{iHt} A e^{-iHt} B} = \frac{1}{Z} \Tr{e^{iH(t+i \beta)} A e^{-iHt} B} \\
        &= \frac{1}{Z} \Tr{e^{iH(t+i \beta)} A e^{-iH(t+i \beta)} e^{-\beta H} B} = \frac{1}{Z} \Tr{e^{-\beta H} B e^{iH(t+i \beta)} A e^{-iH(t+i \beta)}} \\
        &= \Tr{\rho_{\beta} B e^{iH(t+i \beta)} A e^{-iH(t+i \beta)}} = \langle B A(t+i \beta) \rangle_{\beta}.
    \end{split}
\end{equation}

Tomita-Takesaki theorem tells us that we can indeed generalize the result found for qubits to generic physical systems; namely, a (fully) entangled state always gives rise to a dynamics with respect to with an observer, described in terms of its observables, perceives the state as in thermal equilibrium.

Consider a von Neumann algebra $\mathcal{A}$ with a cyclic and separating vector $\ket{\omega}$. Calling $S_{\omega}$ the closure of the map $A \ket{\omega} \rightarrow A^{\dagger} \ket{\omega}$, its polar decomposition can be written as
    
    \begin{equation}
        S_{\Omega} = J_{\omega} \sqrt{\Delta_{\omega}} \text{,}
    \end{equation}

where $J_{\omega}$, called \textit{modular conjugation}, is an anti-isometry and $\Delta_{\omega} = S_{\omega}^{\dagger} S_{\omega}$ is a positive self-adjoint operator, called \textit{modular operator}. Tomita-Takesaki theorem states that the relations
    
    \begin{equation}
        \begin{cases}
            &\Delta_{\omega}^{-is} \mathcal{A} \Delta_{\omega}^{is} = \mathcal{A} \\
            &J_{\omega} \mathcal{A} J_{\omega} = \mathcal{A}'
        \end{cases}
        \label{eq:TT}
    \end{equation}

hold. The set of automorphisms $\{ \sigma_s^{\omega} (\cdot) \}_{s \in \mathbb{R}} \equiv \{ \Delta_{\omega}^{-is} (\cdot) \Delta_{\omega}^{is} \}_{s \in \mathbb{R}}$ is called \textit{modular automorphism group}. Defining the \textit{modular Hamiltonian} $H_{\omega}= - \log{(\Delta_{\omega})}$, the first of \ref{eq:TT} resembles a Heisenberg time-evolution with $s$ being a sort of adimensional time \cite{Longo:2019agc}. Moreover, the modular group satisfies the KMS condition

\begin{equation}
    \bra{\omega} \sigma_s^{\omega}(A) B \ket{\omega} = \bra{\omega} B \sigma_{s+i}^{\omega}(A) \ket{\omega}, \ \forall A, B \in \mathcal{A}
\end{equation}

at inverse temperature $1$. Defining $H_{\text{mod}} \equiv \beta^{-1} H_{\omega}$ and $t \equiv \beta s$, we recover the dimensionful thermal property for $U(t) \equiv e^{iH_{\text{mod}}t}$. The Unruh effect can be derived in this framework by deducing the modular Hamiltonian corresponding to a Rindler wedge \cite{Bisognano:1975ih}, where $J_{\omega} = \mathsf {CPT R}_x(\pi)$ and $\Delta_{\omega} = e^{-2 \pi K}$.

\section{Mellin transform and strip of convergence}

The Mellin transform of a function $\psi$ belongs to the class of transforms we can define on suitable functional spaces. In analogy to Fourier or Laplace transforms, the Mellin transform is defined in integral form:

\begin{equation}
    (\mathcal{M} \psi)(s) = \frac{1}{\sqrt{2 \pi}} \int_0^{\infty} \frac{dk}{k} k^s \psi(k)
\end{equation}

In general, the Mellin transform is defined for complex $s = \sigma + i \omega$. The Mellin integral is convergent in a vertical strip $a < \sigma < b$, with $a$ and $b$ depending on the asymptotic behavior of the function $\psi$ of which we are performing the transform. For instance, if

\begin{equation}
    \begin{cases}
        \psi(k) &\xrightarrow{x \rightarrow 0^+} O(k^{-\alpha}) \\
        \psi(k) &\xrightarrow{x \rightarrow \infty} O(k^{-\beta})
    \end{cases}
\end{equation}

we have $(a,b)=(\alpha, \beta)$. This means that $\omega$ may be analytically continued in the imaginary direction for $-b < \mathfrak{Im}(\omega) < -a$, because $\sigma = - \mathfrak{Im}(\omega)$. Notice that for $\psi(k) = e^{\pm ikv}$ we have that

\begin{equation}
    \int_0^{\infty} \frac{dk}{k} k^s e^{\pm ikv}
\end{equation}

converges for $0 < \sigma < 1$, since near $0$ the phase becomes $1$ and near infinity we need a decreasing function multiplying the rapidly oscillating exponential, by Dirichlet test. The \textit{fundamental strip}, which is the largest strip for which the Mellin transform converges, is given by the interval $(0,1)$. If $\psi(k) = f(k) e^{\pm ikv}$ and $f$ is normalized according to the measure $dk/k$, then

\begin{equation}
    \begin{cases}
        f(k) &\xrightarrow{x \rightarrow 0^+} O(k^{\varepsilon_1}) \\
        f(k) &\xrightarrow{x \rightarrow \infty} O(k^{-\varepsilon_2})
    \end{cases}
\end{equation}

for any $\varepsilon_1, \varepsilon_2 > 0$. As a consequence, we can analytically continue to $- \varepsilon_1 < \sigma < 1 + \varepsilon_2$. The fundamental strip is given by $\bigcap_{\varepsilon_1, \varepsilon_2 \in \mathbb{R}^+} (- \varepsilon_1, 1 + \varepsilon_2 ) \times \mathbb{R} = (0,1) \times \mathbb{R} \subset \mathbb{C}$.

\end{document}